\documentclass[12pt]{article}
\usepackage{amssymb,amsmath,esint,titlesec,mathrsfs}
\titleformat*{\section}{\normalsize\bf}
\titleformat*{\subsection}{\small\bf}

\begin{document}


\begin{titlepage}

\setlength{\baselineskip}{18pt}

                               \vspace*{0mm}

                             \begin{center}

{\LARGE\bf Riemannian \  submersions \  for \  $\mathbf{q}$-entropies }

                                   \vspace{40mm}

              \large\sf  NIKOLAOS \  KALOGEROPOULOS $^\dagger$\\

                            \vspace{1mm}

  \normalsize\sf  Center for Research and Applications \\
                                  of Nonlinear Systems  \ \  (CRANS),\\
                          University of Patras, Patras 26500, Greece.\\

                            \vspace{2mm}
                         
                                    \end{center}

                            \vspace{20mm}

                     \centerline{\normalsize\bf Abstract}
                     
                           \vspace{3mm}
                     
\normalsize\rm\setlength{\baselineskip}{18pt} 

In an attempt to find the dynamical foundations for $q$-entropies, we examine the special case of Lagrangian/Hamiltonian systems of many degrees of freedom whose statistical behavior is conjecturally 
described by the $q$-entropic functionals. We follow the spirit of the canonical ensemble approach. 
We consider the system under study as embedded in a far larger total system. We explore some of the consequences that such an embedding has, if it is modelled by a Riemannian submersion. 
We point out the significance in such a description of the finite-dimensional Bakry-\'{E}mery Ricci tensor, 
as a local mesoscopic invariant, for understanding the collective dynamical behavior of systems described by the $q$-entropies.    

                           \vfill

                                                                         
                             \vfill

\noindent\rule{9cm}{0.3mm}\\  
   \noindent   $^\dagger$ {\normalsize\rm Electronic mail: \ \ \ \  {\sf nikos.physikos@gmail.com}}\\

\end{titlepage}
 

                                                                                \newpage                 

\rm\normalsize
\setlength{\baselineskip}{18pt}

\section{Introduction}

The central goal  of Statistical Mechanics has been the derivation, as much as possible, of the behavior of
macroscopic systems from their microscopic constituents. Entropy has  played a central role in this quest, 
during the last century and a half, since its introduction by R. Clausius and its probabilistic interpretation by L. Boltzmann and G.W. Gibbs. 
Despite the considerable progress that has been made in the intervening years, it is probably fair to say that there are many aspects of entropy which are 
not very well-understood, not only at the conceptual or epistemological levels, but even more so in our attempts to calculate the macroscopic  from the 
microscopic dynamics of the systems under study \cite{GLTZ}.  \\

The advent of the $q$-entropy \cite{Tsallis}, as well as of a whole host of entropic functionals that have been proposed in Physics since that time \cite{IKGS}, 
has provided another incentive to look closer into the dynamical foundations of entropy, and therefore indirectly of temperature. 
Even though a lot of work has been dedicated toward understanding the thermodynamic 
implications \cite{T-book} of the $q$-entropy, very little is actually known about its dynamical foundations. 
As a by-product of such a study which we are pursuing in the present work,  we expect to be able to get a clearer understanding the dynamical 
foundations of the Boltzmann and Gibbs functionals which are used, almost exclusively and canonically, in Statistical Mechanics.\\

In this work, we make yet another  attempt in the direction of understanding the dynamical foundations of entropy.
We start from an isolated system, one could call it the ``Universe", a word which is not meant to be taken literally. 
We model the system under study as a subsystem of such a ``Universe". We follow the standard ``canonical ensemble" approach of J.W. Gibbs.
We couple the system under study to its environment, which is far bigger. The union of the system under study and the environment is the ``Universe".
We employ a geometric treatment, expressing the dynamics of system under study as a Riemannian submersion of the ``Universe". 
The subset of the points of the ``Universe" which is mapped into the same point in phase space of the system under study is interpreted as a very large 
thermostat. Such thermostats are in contact with the system and can be the sources of different temperatures during the evolution of the system. 
As a result, this formalism is sufficiently flexible to allow for a description of non equilibrium systems. \\

The result, under some additional assumptions clearly spelled out in this work, is that the collective dynamics of the  underlying system 
is provided by the finite dimensional  Bakry-\'{E}mery(-Qian) Ricci tensor.  The relation of this tensor to the $q$-entropy is given 
by results produced  in the line of works culminating with  \cite{S1, S2, LV} and continuing until  today \cite{Ohta, Ambrosio}.\\    

In Section 2, we present some details about Riemannian submersions and the role that they have in our treatment. 
In Section 3, we explain how Riemannian submersions fit within the context of smooth metric measure spaces, which is a natural geometric framework
for the $q$-entropy and point out the role of the generalized Barky-\'{E}mery(-Qian) Ricci tensor and its lower bounds. 
Section 4, makes some further comments and presents open issues that may be worth exploring in the future.  \\   


\section{A ``canonical ensemble"-like approach via Riemannian submersions}

We start with a system, the ``Universe"  having a large, but finite, number of degrees of freedom. Let \ $\mathcal{M}$ \ be its configuration or phase space, which is assumed 
to be a $N$-dimensional manifold, where $N$ is a, finite, positive integer. In the case that \ $\mathcal{M}$ \ is the phase space, obviously $N$ should be even. 
One can allow \ $\mathcal{M}$ \ to have some mild form of singularities, but this would not impact in any obvious way the essence of the arguments which follow.  \\

Let us assume that we are only interested in analyzing point particle models, so we ignore the existence of fields in this work.
Fields are maps, usually sections of vector or spinor bundles with connections over spacetime. 
As such, they give rise to infinite dimensional configuration and phase spaces. 
Even a mathematically simplistic treatment of such structures is quite non-trivial and presents difficulties,  
which we prefer to avoid, for the sake of simplicity,  in the present work. \\

Part of the ``Universe" is the system that we choose to study. It is a system of  fewer degrees of freedom than the ``Universe", which is thermally coupled to it. 
We imagine \ $\mathcal{B}$ \ as being the configuration or the phase space of the system under study.  
In this work, we consider a meso- or macro-scopic description of \ $\mathcal{B}$, \  in the sense of describing \ $\mathcal{B}$ \ after some coarse-graining has taken place.
Coarse-graining is a fundamental aspect in the description of systems of many degrees of freedom where entropy plays a fundamental role \cite{Gor, Kal1, Kal2}. \\

We proceed in analyzing the system by using a geometric approach. To this end, we choose a Riemannian metric describing such a system. 
Since, in many cases of physical interest, the microscopic particle Hamiltonian has the form   
      \begin{equation}
              \mathscr{H}(q_i, p_i) \ = \ \sum_{i=1} ^{3N} \frac{p_i^2}{2m_i} + \mathscr{V} (q_1, q_2, \ldots, q_{3N})
      \end{equation}
where \ $\mathscr{V}(q_1, q_2, \ldots q_{3N})$ \ stands for the position-dependent potential energy of the system. 
The parameters \ $m_i$ \ can be considered as effective masses of the corresponding degrees of freedom which can be set to one,
by rescaling. A Riemannian metric can be chosen as an extension of the quadratic term containing the canonical momenta. 
In the above case, the metric can be chosen to be the diagonal Euclidean one, whose components are 
\begin{equation}  
        g_{ij} \ = \  \delta_{ij}, \hspace{15mm} i, j = 1, \ldots, 6N
\end{equation}
in the coordinate basis \ $(\frac{\partial}{\partial q^i}, \frac{\partial}{\partial p_i}), \ \  i=1,\ldots, 3N$ \ associated to the coordinate system \ $(q_i, p_i)$. \
In a Lagrangian description, and for a system with total conserved energy \ $\mathscr{E}$, \ a common choice is the Jacobi metric \cite{Ong}
\begin{equation}
          \tilde{g}_{ij} \ = \  2[\mathscr{E}-\mathscr{V}(q_1, \ldots, q_{3N})] \ g_{ij}, \hspace{15mm} i, j = 1, \ldots, 3N
\end{equation} 
where \ $g_{ij}$ \ is derived from the quadratic/kinetic energy  term of the particle Lagrangian in configuration space.  \\

It is unclear,  what would be a ``best" or ``most appropriate" choice of such a metric, or even if such adjectives make any sense beyond satisfying the criterion of the convenience 
of explicit calculations in such physical models. It is evident that there considerable level of arbitrariness in choosing such  metrics and someone would have to eventually 
ascertain that the experimentally measurable quantities of the model do not depend on such choices.\\

There are many ways to model geometrically the interaction of the ``Universe"/$\mathcal{M}$ and the system under study/$\mathcal{B}$. 
One such way, which may prove to be particularly fruitful is as a Riemannian submersion. For simplicity let us assume throughout this 
treatment that both \ $\mathcal{M}$ \ and \ $\mathcal{B}$ \ are  smooth, namely they belong to the \ $C^\infty$ \ class. 
We call \ $\mathcal{B}$ \ the base.
A submersion \ $\pi$ \ is, in a sense, a concept dual to that of an immersion. 
Consider a \ $C^\infty$ \ map \ $ \pi : \mathcal{B}\rightarrow\mathcal{B}$ \ such that  at each point 
\ $x\in\mathcal{M}$, \ its differential
\begin{equation}  
        d\pi_x: \ T_x\mathcal{M} \ \longrightarrow \ T_{\pi(x)}\mathcal{B}
\end{equation}
is a surjective map. If this property holds for all \ $x\in\mathcal{M}$ \ then \ $\pi$ \ is called a submersion of \ $\mathcal{M}$ \ onto \ $\mathcal{B}$. \  
In case  such a submersion is also a proper map, namely if inverse images of  compact subsets are compact, then according to C. Ehresmann's lemma 
such a map is a locally trivial fibration. A fibration is almost a fiber bundle, if the inverse images of all points of the base \ $\mathcal{B}$ \ have inverse images of 
the same dimension, which  will be the case of interest in this work. \\

We use the concept of submersion to encode the physically reasonable fact that  any part of the system under study,
modelled by \ $\mathcal{B}$, \ is the result/projection of some subsystem of the ``Universe", as modelled by \ $\mathcal{M}$, \ and that there is no part
of \ $\mathcal{B}$ \ which does not arise as such a ``projection" of some part of \ $\mathcal{M}$. \\  

All of this above takes place at the level of smooth structures. However, as has been mentioned above, we will also assume the existence of a metric 
structure \ $g$ \ on  \ $\mathcal{M}$, \ which may not necessarily be unique. Due to the choice of \ $g$ \ one has an orthogonal decomposition of the tangent space
at any point \ $z\in\mathcal{M}$ \ given by
\begin{equation} 
       T_z\mathcal{M} \ = \ \mathcal{V}T_z\mathcal{M} \oplus \mathcal{H}T_z\mathcal{M} 
\end{equation}
where \ $\mathcal{V}T_z\mathcal{M}$ \ denotes the vertical part of the tangent space, 
made up of all elements of \ $T\mathcal{M}$ \ which are  tangent to the fibers \ $\mathcal{F}_z = \pi^{-1}(\pi(z))$ \ and 
\ $\mathcal{H}T_z\mathcal{M}$ \ denotes the horizontal part of the tangent space, namely the orthogonal complement of \ 
$\mathcal{V}T_z\mathcal{M}$ \ in \ $T_z\mathcal{M}$ \ with respect to \ $g$. \\
   
If, in addition, the submersion \ $\pi$ \ preserves the length of the horizontal vectors, namely if the map
\begin{equation}
   d\pi_z: \ \mathcal{H}T\mathcal{M} \ \longrightarrow \ T_{\pi(z)}\mathcal{B}
\end{equation} 
is an isometry, then \ $\pi$ \ is called a Riemannian submersion. The systematic study of Riemannian submersions started with \cite{Nag} and reached its 
full development in the works \cite{ON1, ON2, Gray}. Many things have been known about Riemannian submersions since these foundational works \cite{Besse, GLP, FPI, GW}. 
An example is  the following theorem of R. Hermann : any Riemannian submersion \ $p: \mathcal{M} \longrightarrow \mathcal{B}$ \ where \ $\mathcal{M}$ \  is assumed to be complete,
is a locally trivial fiber bundle \cite{Hermann}. If \ $y\in\mathcal{B}$, \ then the implicit function theorem implies that the inverse image of \ $y$, \  
namely the fiber \ $\pi^{-1}(y)$ \ is a closed sub-manifold of \ $\mathcal{M}$ \  whose dimension is 
\begin{equation}
          \mathsf{dim}\mathcal{F} = \mathsf{dim}\mathcal{M} - \mathsf{dim}\mathcal{B}
\end{equation}
Research on Riemannian submersions and their metric generalizations called ``submetries" \cite{Berest}  continues to this day \cite{Lempert}. 
It may be worth  noticing that even though the vertical distribution \ $\mathcal{V}T\mathcal{M}$ \ is integrable, being tangent to the fiber \ $\mathcal{F}$, \  
this is generally not the case for the horizontal one \ $\mathcal{H}T\mathcal{M}$. \\    

To make things more concrete, B. O'Neill defined in \cite{ON1}, the following two tensors, for \ $E, F \in T\mathcal{M}$, \ where \ $\nabla$ \ is the Levi-Civita 
connection associated to the metric \ $g$ \ on \ $\mathcal{M}$: 
\begin{equation}  
   \mathbb{T}_E F \ = \ \mathcal{H}\nabla_{\mathcal{V}E}(\mathcal{V}F)  + \mathcal{V} \nabla_{\mathcal{V}E}(\mathcal{H}F)
\end{equation}
Throughout this work \ $\mathcal{V}$ \ and \ $\mathcal{H}$ \ indicate the vertical and horizontal projections of the corresponding vectors, respectively. 
Many properties of \ $\mathbb{T}$ \ have been derived in \cite{ON1}. For our purposes, it is sufficient to know that \ $\mathbb{T}$ \ is the second fundamental form of the 
induced metric on the fiber \ $\mathcal{F}$ \ if we restrict our attention to vertical vector fields only. As a result, \ $\mathbb{T} = 0$ \ is equivalent to stating  that the fibers 
\ $\mathcal{F}$ \ are  totally geodesic sub-manifolds of \ $\mathcal{M}$.  \\

The second important tensor defined in \cite{ON1} for Riemannian submersions 
is the integrability tensor \ $\mathbb{A}$.\ It is constructed by interchanging \  $\mathcal{H}\leftrightarrow\mathcal{V}$ \ in the definition of \ $\mathbb{T}$: 
\begin{equation}
            \mathbb{A}_E F \ = \  \mathcal{H}\nabla_{\mathcal{H}E}(\mathcal{V}F) + \mathcal{V}  \nabla_{\mathcal{H}E}(\mathcal{H}F)
\end{equation}
The characterization of \ $\mathbb{A}$ \ as ``integrability tensor" stems from the fact that for \ $X,Y \in \mathcal{H}T\mathcal{M}$ \ one \cite{ON1} can prove that
\begin{equation}   
      \mathbb{A}_X Y \ = \ \frac{1}{2} \ \mathcal{V} ([X,Y])  
\end{equation}
where \ $[\cdot, \cdot]$ \ stands for the Lie bracket of the corresponding vector fields.  
As a result, \ $\mathbb{A} = 0$ \ is equivalent to stating that the horizontal distribution of the submersion is integrable. \\

In the sequel we will assume that \ $\mathcal{M}$ \ is complete with respect to \ $g$. \  This is a reasonable assumption for the particle models we are intersted in analyzing. 
Let \ $\gamma(t): [0,1] \longrightarrow \mathcal{B}$ \ be a smooth curve on \ 
$\mathcal{B}$. \ Let \ $\mathcal{F}_{\gamma(0)} = \pi^{-1}(\gamma(0))$ \ and \ $\mathcal{F}_{\gamma(1)} = \pi^{-1}(\gamma(1))$ \
be the two fibers over the endpoints of \ $\gamma$. \ Choose a point \ $a\in \mathcal{F}_{\gamma(0)}$ \ and let \ $\widetilde{\gamma}$ \ be the horizontal lift of \ 
$\gamma$ \ on $\mathcal{M}$ \ having \ $a$ \ as its starting point. Let \ $\varphi(a)$ \ be the endpoint \ $\widetilde{\gamma}(1)$. \ Then \ $\varphi$ \ is a  
fiber diffeomorphism from \ $\mathcal{F}_{\gamma(0)}$ \ to \ $\mathcal{F}_{\gamma(1)}$. \ The necessary and sufficient condition for the diffeomorphism \ $\varphi$ \ 
to be an isometry between the fibers, is that the fibers are totally geodesic \cite{ON1}. Moreover, if \ $\widetilde{\gamma}$ \ is a geodesic on \ $\mathcal{M}$ \ which is 
horizontal at \ $a$, \ then \ $\gamma$ \ is also a geodesic of \ $\mathcal{B}$ \cite{ON2}.   \\

A physical interpretation of some of the above geometric concepts in the current context are as follows. The fibers \ $\mathcal{F}$ \ of the submersion \ $\pi$ \ represent the 
``environment", at each time instant during the evolution of the system under study. The fibers are the degrees of freedom over which one integrates if they wish to get an effective 
description of the dynamics of the system under study.  In the canonical ensemble approach, \ $\mathcal{F}$ \ represent the thermostats with which the system \ $\mathcal{B}$ \  
is in contact as it evolves. The present framework allows for thermostats having different temperatures during the evolution of the system. 
Different temperatures can be used to drive or keep  the system out of equilibrium. \\

An important question is whether  ``temperature" is, or can be represented as a geometric quantity in the present context.
Given the multitude of different definitions of temperature \cite{Biro, HHD} and the lack of consensus about which one would be appropriate and 
measurable in each experiment, the answer to the above question is unclear. So one could insist that temperature is a quantity that can be encoded geometrically 
in the proposed framework of Riemannian submersions. The question is then what exactly is the geometric representation of temperature. \\

The immediate thought is to choose between two of the simplest geometric quantities of the ``thermostat" \ $\mathcal{F}$ 
its diameter or its volume. We want to have the possibility of allowing for ``infinite" thermostats, at least in an appropriate limit. For this reason, we exclude the diameter 
as a possible indicator of temperature. Therefore, we are left with using the volume of \ $\mathcal{F}$ \ as a simple geometric quantity representing the temperature of the 
thermostat \ $\mathcal{F}$. \\         
       
 The above choice can bring forth some criticisms. On technical grounds, we prefer to use a fiber \ $\mathcal{F}$ \ which is compact. This is not a problem in 
 a Lagrangian description, especially if the system is spatially confined, but can become problematic in a Hamilltonian description. For such cases, one can initially put the system in
 a ``box", whose sides are used as regulating quantities, and at the end of the calculation of the quantities of interest  take the limit  as the sides of this box grow to infinity. 
 This is a rather common approach in several parts of Physics, but it should be done carefully and can be hard to justify at times, depending on the properties of the system under study. 
 It is especally problematic for systems with long-range interactions that the $q$-entropies aim to describe.\\
 
 If one considers an ideal gas, then the temperature is related to the average kinetic energy of its particles, which in turn depends on the average speed of the particles. 
 But the speed depends on ``time" which is clearly not a geometric quantity. Therefore it appears that associating the temperature to a geometric quantity may not a good idea. \\
 
 A way to reply to such criticism is by invoking the local ergodicity of the system under study. 
Practically very hard to prove, or even argue about in a convincing way for concrete systems, 
ergodicity of the evolution of the microscopic system under study  is invoked at times, when one studies the foundations of Statistical Mechanics. Roughly speaking, 
the idea is to replace time averages with averages over subsets of the space on which the system evolves. One may not be able to argue that a system is ergodic, but 
it may be easier for someone to accept local ergodicity. This in the spirit of localizing a global property in order to gain some level of generality and flexibility in the 
underlying mathematical structure.\\

Objections to local ergodicity may occur in Statistical Mechanics which stem from to the enormity of the 
Poincar\`{e} recurrence time \cite{KH}, as opposed to the  finite (macroscopic) relaxation time to equilibrium etc. But one should remember that sevela of the microscopic
features of the system have already been ignored, since we are working at the level of a meso- or macro-scopic ``coarse-grained", rather than a microscopic ``fine-grained" 
description of the system. To strengthen the argument about the validity of assuming local ergodicity, one can invoke  the ergodic decomposition theorem \cite{KH} which
states, very roughly,  that any measure preserved by a flow can be written as a sum of ergodic measures. Hence the ergodic measures are a ``basis" in terms of which
we can decompose any flow-preserved measures, such as the phase space volume that we see in a Hamiltonian description of the system. \\

So, if we follow this line of reasoning, even though the temperature of one system may depend on time and thus not be considered geometric, the overall effect of 
studying an ensemble of such systems may eliminate such a dependence. And since we want to allow for systems out of equilibrium all that we need is local, rather than global, 
ergodicity, since we do not wish to necessarily explore the limit of the ``long time" behavior of the system. From a dynamical systems viewpoint exploring the ``finite time" 
behavior of the system under study would be  a momentous task, as appropriate quantities would have to be defined or existing quantities would have to be modified  ad calculated 
to attain such a goal.\\

An obvious question is why use Riemannian submersions? The answer is that these are structures which are flexible enough to accommodate new features, but at the same 
time they preserve features of more familiar structures. A Riemannian submersion is a more general concept than that of a projection, something which is desirable, as
projections are associated with marginal probability distributions which are used extensively in Statistical Mechanics. On the other hand, Riemannian submersions are not
``too flexible". \\

The requirement that the map (6) be an isometry is quite reasonable: there is no need to assume that the horizontal distributions are not isometric to that of the
base \ $\mathcal{B}$ \ since we plan to associate most aspects of the coupling of the system under study with its environment to the fibers \ $\mathcal{F}$ \ of the submersion. 
Giving up such an isometry of horizontal spaces would open up far too many possibilities, which would be hard to control. This is 
a major reason why we believe that Riemannian submersions are a geometric structure which strikes a reasonable balance between generality and controllable behavior, 
and thus are a good choice for reaching our goals.\\         


\section{Riemannian submersions for smooth metric measure spaces}

We presented in \cite{Kal3} a simplified and pedestrian argument whose goal was to show how the convexity of the $q$-entropy is related to the lower bounds of a generalized Ricci 
curvature on a metric measure space. This argument is established in full generality and rigor and is presented, not only for $q$-entropies, in \cite{S1, S2, LV}. 
For an extensive background on this matter, one may also wish to consult \cite{V-book} and for some more recent developments, in the spirit of gradient flows \cite{Ambrosio}. \\

In the present work we will focus on a ``canonical ensemble"/Gibbsian treatment of aspects of this line of development, 
focusing exclusively on Riemannian submersions of smooth metric measure spaces and their generalized Bakry-\'{E}mery(-Qian) Ricci tensor. We follow closely the 
results of \cite{L1, L2} and only add some comments of potential interest to Physics. A relatively recent review focusing on results of comparison geometry needed in 
the present work, can be found in \cite{WW}. For the definitions, meaning and mathematical significance of the (measured-) Gromov-Hausdorff convergence, to which we allude in 
the present work, one may wish to consult \cite{LV, V-book}, or the foundational \cite{Gr-book}.  \\  

For our purposes it will be sufficient to assume that the ``Universe" whose configuration/phase space is \ $\mathcal{M}$ \ is a Riemannian manifold. However, the formalism has been developed
 for the more general case where \ $\mathcal{M}$ \ is a smooth metric measure space. A smooth metric measure space is a Riemannian manifold \ $(M, g)$ \ endowed with a measure
 \begin{equation}
 d\mu \ = \  \phi \ d\mathsf{vol}_\mathcal{M}  
 \end{equation}
 where \ $d\mathsf{vol}_\mathcal{M}$ \ is the infinitesimal  Riemannian volume element of \ $\mathcal{M}$.  \ We assume that \ $\phi : \mathcal{M} \rightarrow \mathbb{R}_+$ \ is a positive and 
 sufficiently smooth function on \ $\mathcal{M}$. \ On physical grounds, we will have to assume that \ $\phi$ \ should be a sufficiently simple function which can depend on the particular model at hand.
 The most obvious choice is to consider \ $\phi$ \ to be constant and equal to one on \ $\mathcal{M}$, \ in which case we return to \ $\mathcal{M}$ \ being an $N$-dimensional Riemannian manifold.  \\ 
 
The Riemann tensor of a smooth metric measure space is independent of the measure, so its definition remains the same as that for a Riemannian manifold. On the other hand a generalization of the 
Ricci and scalar tensors is called for, since they both depend on the measure \ $d\mu$. \ A generalization of the Ricci tensor on a smooth metric measure space \ $\mathcal{M}$ \ is not unique. Let 
\  $\mathrm{Ric}$ \ denote the Ricci tensor of the $N$-dimensional Riemannian manifold \ $(\mathcal{M}, g)$. \
After a lot of work spanning almost two decades, it was found that the following  Bakry-\'{E}mery(-Qian) generalized ($N$-) Ricci tensor has many desirable properties:
 \begin{equation}
        \overline{\mathrm{Ric}}_\infty \ = \ \mathrm{Ric} - \mathrm{Hess} (\log \phi)
 \end{equation}
 and 
 \begin{equation}
         \overline{\mathrm{Ric}}_N \ = \ \mathrm{Ric} - \mathrm{Hess} (\log \phi) - \frac{1}{N-n} \ d\log\phi \otimes d\log\phi
 \end{equation}
 which can be rewritten as 
 \begin{equation}
         \overline{\mathrm{Ric}}_N \ = \ \mathrm{Ric} - (N-n) \  \frac{\mathrm{Hess (\phi^\frac{1}{N-n})}}{\phi^\frac{1}{N-n}}
 \end{equation}
 In the above definitions, in the  context of Riemannian submersions, $n$ denotes the dimension of the base manifold \ $\mathcal{B}$ \ so that \ $N-n \in [0, +\infty]$ \ is the dimension of the fibers 
 \ $\mathcal{F}$ \ according to (7). Moreover, \ $\mathrm{Hess}(\cdot )$ \ stands for the Hessian function of its argument.\\ 
 
  We denote, in passing, that these definitions of the generalized Ricci tensor of smooth metric measure spaces 
 are independent of the context of submersions, in their full generality \cite{S1, S2, LV}. 
 In the most general context of smooth metric measure spaces \ $N$ \ is a form of a ``dimension" of the measure \ $d\mu$ \ and \ $n$ \ is the dimension of the underlying
 Riemannian manifold. For this reason, in general, \ $N$ \ does not have to be an integer; all it has to be is greater than \ $n$. \ 
 Since we are working in a meso- or macro- scopic level which has resulted after some form of coarse-graining has taken place over the microscopic dynamics, 
 both \ $N$ \ and \ $n$ \  are far smaller than the \ $10^{23}$ \ dimension of the configuration or phase space of the microscopic system. 
 The rest of this work tacitly assumes the content of these comments.\\
 
 The tensor (12) was  initially proposed for infinite \ $N$ \ by Bakry and \'{E}mery \cite{BE} and was subsequently extended to (13) and investigated for finite \ $N$ \ by Qian \cite{Qian}. 
 Its further significance for geometry, in the context of of optimal transportation, which does not assume smoothness of the underlying manifold \ $\mathcal{M}$, \ 
 its indirect relation to the $q$-entropies through the \ $DC_N$ \ displacement convexity classes, 
 its stability properties under measured Gromov-Hausdorff convergence and further geometric implications culminated with the work of \cite{S1, S2, LV}.  \\
 
 To continue, let us assume that the fibers of the Riemannian submersion \ $\pi: \mathcal{M} \rightarrow \mathcal{B}$ \ are compact, for the reasons indicated at the end of Section 2. 
 Let \ $\overline{f}$ \ be a positive function on \ $\mathcal{M}$ \ and define the smooth positive function \ $f$ \   on \ $\mathcal{B}$ \ by  
 \begin{equation} 
     \pi_{\ast} (\overline{f} \ d\mathsf{vol}_\mathcal{M}) \ = \ f \ d\mathsf{vol}_\mathcal{B}
 \end{equation}    
Using the notation of Section 2, J. Lott proved in \cite{L1} the following. Assume that the fiber transport map \ $\varphi$ \ preserves the fiber measure \ $f \ d\mathsf{vol}_\mathcal{F}$ \ 
up to a multiplicative constant, namely if for any smooth curve \ $\gamma: [0,1] \rightarrow \mathcal{B}$ \ there is a constant \ $c(\gamma) > 0$ \ such that 
\begin{equation}     
     \varphi^\ast \left(\overline{f}\big|_{\mathcal{F}_{\gamma(1)}} \ d\mathsf{vol}_{\mathcal{F}_{\gamma(1)}} \right) \ = \ c(\gamma) \  \overline{f} \big|_{\mathcal{F}_{\gamma(0)}} \ 
                                                                                                                           d\mathsf{vol}_{\mathcal{F}_{\gamma(0)}}  
\end{equation}
then if on \ $\mathcal{M}$
\begin{equation}
          (\overline{\mathrm{Ric}}_\mathcal{M})_\infty \ \geq \ C \ g_\mathcal{M}   
\end{equation}
the same applies for the corresponding tensors on \ $B$, \ where \ $C\in\mathbb{R}$ \ is a constant. More compactly, the theorem states that
\begin{equation}
            (\overline{\mathrm{Ric}}_\mathcal{B})_\infty \ \geq \ C \ g_\mathcal{B}       
\end{equation}
In short, a preservation of the measure of each fiber up to a multiplicative constant
by the Riemannian submersion on the smooth metric measure space, implies an O'Neill type relation for the lower bound of the Ricci curvature.  \\

We may wish to recall at this point, that one of the most useful results of \cite{ON1} is the fact that the sectional curvature 
increases under Riemannian submersions. This has provided a way of explicitly constructing manifolds of positive sectional curvature.  
Moreover, under the additional assumption that    
\ $\overline{f} = 1$, \ it was proved in \cite{L1} that if on \ $\mathcal{M}$ \ the ordinary   Ricci tensor has a lower bound
\begin{equation}
   \mathrm{Ric}_\mathcal{M} \ \geq \ C \ g_\mathcal{M}
\end{equation} 
then on the base of the submersion \ $\mathcal{B}$ \ the generalized Bakry-\'{E}mery(-Qian) $N$-Ricci tensor has the same lower bound  
\begin{equation}
   (\overline{\mathrm{Ric}}_\mathcal{B})_{N-n} \ \geq \ C \ g_\mathcal{B}
\end{equation}
where we have used explicitly the subscripts to indicate which tensors are referring to which manifolds. \\  

The significance, for Statistical Mechanics, of the above result contained in (19), (20)  is the following: First of all, we do have to assume that the Ricci curvature of 
the coarse-grained configuration or phase space \ $\mathcal{M}$ \  of the ``Universe" has a lower bound. If not, then due to the Bishop-Gromov inequality \cite{Gr-book}, 
the volume element may  increase without bounds, something which would be troubling physically and hard to control mathematically. It happens, for instance, for 
the case of the Jacobi metric (3) as pointed out in \cite{Ong} and this has created problems or discourages someone from using this metric at partially or at all, but 
especally close to the Hill boundary of the configuration space.\\

Preserving up to a overall constant the fiber measure under
fiberwise diffeomorphisms is a rather strong condition, which however is quite a bit weaker than demanding that all fibers be isometric. This happens in the case where one 
chooses a submersion such that the second fundamental form \ $\mathbb{T}$ \ in (8) were zero, so the corresponding fibers are totally geodesic submanifolds of \ 
\ $\mathcal{M}$. \  This assumption would only allow for thermostats having the same temperature, which signifies not a very flexible condition. 
By contrast, the preservation condition on the fiberwise   measure up to  multiplicative constant 
allows us the flexibility of using the volume of each fiber as being related to the temperature of a thermostat and still allowing for the system to be coupled to thermostats of different temperatures, 
thus allowing us to deal with non-equilibrium systems. \\

Setting the function \ $\overline{f} = 1$ \ is a matter of simplicity. We have no obvious reason to make any 
other choice for \ $\overline{f}$. \ One could start from a well-understood system as the ``Universe", and model it on \ $\mathcal{M}$ \ by demanding the underlying Hamiltonian 
system  to be uniformly hyperbolic or ergodic \cite{KH}, such as the geodesic flow on a negatively curved manifold, for instance.  Then, by choosing a Lagrangian or Hamiltonian, 
we can effectively set up a Riemannian submersion and proceed by analyzing the coarse-grained system using geometric methods.    \\

The theorem contained in (19), (20) also provides a dynamical basis for the $q$-entropies. As was stated in \cite{Kal3}, following \cite{S1, S2, LV}, convexity bounds of the 
$q$-entropy functional on the Wasserstein space of measures, are equivalent to lower bounds on the Bakry-\'{E}mery(-Qian) generalized Ricci tensor \ $\overline{\mathrm{Ric}}$ \ 
of the underlying metric measure space.  It was also pointed out in \cite{Kal3} that the relation between \ $q$ \ and \ $N$ \  is 
\begin{equation}
           q \ = \  \frac{N-1}{N}
\end{equation}
where an isoperimetric interpretation for \ $q$ \ was subsequently provided. 
 Since $q$-entropies possess such convexity properties, it is not unreasonable to demand corresponding lower bounds for \ $\overline{\mathrm{Ric}}$ \ 
on physical grounds. Due to this fact, one could ask for a more straightforward construction of \ $\overline{\mathrm{Ric}}$. \ 
The present approach, which follows the Gibbsian approach to 
equilibrium Statistical Mechanics does that. 
So, in a way, it provides a kind of justification of the statement that the $q$-entropies may be appropriate for describing systems out of equilibrium. \\

However one has to be careful at this point. The general arguments leading to the relation between the convexity properties of the $q$-entropies in the Wassertein space of 
\ $\mathcal{M}$ \ and the lower Bakry-\'{E}mery-(Qian) generalized Ricci curvature bounds on \ $\mathcal{M}$ \ depend on the displacement convexity properties of such functionals \cite{LV, V-book}.
There is an infinity of functionals that someone can construct obeying these properties. Hence $q$-entropies are just one set of functionals singled out by these arguments, but are not unique.   
Additional requirements needed to be met in order for someone to be able to single out the $q$-entropies alone in such constructions. \\

A re-assuring feature is that these constructions, which are based on optimal transportation,  are preserved under the rather ``coarse" measured Gromov-Hausdorff convergence.  This 
allows us to believe that their essential features may be preserved even if one takes into account some of their possible perturbations or may not care about some of their fine details, as 
this whole geometric approach works at the level of coarse-grained structures \ $\mathcal{M}, \ \mathcal{B}$ \ for which it may be meaningless to discuss fine details.\\       

Actually, more than the above result was proved in \cite{L1}. It was also proved that a closed, namely compact and without boundary,  metric measure space \ $(\mathcal{B}, f)$ \ 
whose Bakry-\'{E}mery(-Qian) generalized Ricci curvature has a uniform lower bound \ $C$ \ as in (20), is the measured Gromov-Hausdorff limit of a sequence of $N$-dimensional closed 
Riemannian manifolds whose Ricci curvatures have the same lower bound \ $C$. \ So, in a way, and having in mind the caveats stated above, the $q$-entropies of a system can be seen as 
appropriate limits of the Boltzmann-Gibbs-Shannon entropies of systems inside which they are embedded, and from which they arise through the metric process of ``manifold collapse".
This is a process of ``manifold collapse" with Ricci curvature bounded below, which has attracted very substantial interest during the last four decades.
In some sense, we may wish to also see the present way of thinking as an entropic analogue to the Whitney, Nash or any other embedding theorems familiar from Topology and Geometry \cite{Gromov}. \\  


\section{Discussion and further work}

In this work, we discussed a ``canonical" approach, in the Gibbsian sense of the word, related to the origin of the $q$-entropies. 
We relied on Riemannian submersions and the geometry of smooth metric measure spaces. In particular, we pointed the significance of the   
Bakry-\'{E}mery(-Qian) generalized $(N_)$ Ricci tensor for providing a local, but covariant charaterization of the pertinent aspects of such a submersion. 
This tensor is explicitly computable, at least in principle, and should also be in practice, as long as someone can  find a convenient coordinate system
through which to parametrize the coarse-grained configuration / phase space(s) involved, such as \ $\mathcal{M}, \ \mathcal{B}$ \ as well as explicitly be able to
describe the action of the Riemannian submersion \ $\pi: \mathcal{M} \rightarrow \mathcal{B}$. \ Such an explicit description is clearly model-dependent.    \\ 

To develop further the general theory, avoiding any model-specific details, we could try to speculate on what a reasonable form of the manifolds 
\ $\mathcal{M}, \ \mathcal{B}, \ \mathcal{F}$ \  and the Riemannian submersion \ $\pi$ \ could possibly be. Probably the most reasonable choice would be to consider the fibers \ $\mathcal{F}$ \  
to be totally geodesic submanifolds of \  $\mathcal{M}$, \ namely to have \ $\mathbb{T} = 0$. \ Then all such fibers would be ``independent" of each other in a sense,
since any geodesic in one such fiber would remain in it and would be ``unaware" of its embedding into \ $\mathcal{M}$ \ and about  the existence of all the other fibers of the submersion.
The geometry of such Riemannian submersions has been extensively discussed in \cite{Esc}. The requirement for the Riemannian submersion to have totally geodesic fibers would 
be ideal for modelling a system coupled to thermostats all of which have the same temperature, since all such fibers would be isometric, as has been pointed above.  \\

Another possibility would be to assume that the fibers are minimal submanifolds, namely require just the trace of \ $\mathbb{T}$ \ to vanish, rather than the tensor itself. 
This would be harder to justify on physical grounds even though it might describe quite well situations where the system is couples to different temperature thermostats.   
The flip side of this is that far more mathematical possibilities exist for such Riemannian submersions among which one may have to choose an appropriate one for the model at hand. \\

In a similar vein one could choose, for instance, the tensor \ $\mathbb{T}$ \ to be basic \cite{Bord}, or have any similar properties. In this case there are just too many possibilities which would be best 
explored as the need arises in particular models. As yet another option, weakening the requirement of the submersion having totally geodesic fibers, one might wish to require the 
spectrum of the underlying Laplace-Beltrami operator to have particularly nice properties, 
such as to commute with the submersion on functions. 
For totally geodesic fibers this possibility has been worked out in detail in \cite{BBB} .
One can find a similar treatment in parts of \cite{GLP}. \\      

One might wish to be more general and explore the implications of  Riemannian submersions for the generalized scalar curvature of metric measure spaces. 
The scalar curvature is less understood than the Ricci one 
even for Riemannian manifolds \cite{Gr3}. There is no general consensus of what might constitute a good definition of scalar curvature for a metric measure space. 
For a proposal, see \cite{L2}.  Part of the difficulty  is also related to the fact that we do not exactly know what kind of convergence is suitable for scalar curvature, 
in the place of the measured Gromov-Hausdorff convergence, which is used when dealing with Ricci curvature. \\ 

From a physical viewpoint what this approach is lacking are explicit computations on models of potential physical interest. 
In other words, we do lack specific computations and specific predictions which would concretely substantiate the use of such geometric 
ideas to systems, even conjecturally, described by the $q$-entropies. This is a technical but important difficulty in practice, but not a limitation of the formalism per se. \\

As a final point and as has been mentioned before,  we would certainly like to see the 
extension of this approach to infinite-dimensional context, which would be used to describe field theoretical, rather than particle, models. 
We believe that these are topics worth further investigation in the future.\\


              \vspace{7mm}

\noindent{\bf Acknowledgement:} We would like to thank Professor Anastasios Bountis whose encouragement and continuous support made this work possible.



\end{document}